\documentclass[a4paper]{jpconf}
\usepackage{amsmath, amssymb}
\usepackage{amsfonts}
\usepackage{xcolor}
\usepackage{graphicx}
\usepackage{rotating}
\usepackage{colortbl}
\usepackage{subfig}
\usepackage{cite}

\newcommand{\ee}{\mathrm{e}}

\newcommand{\bfmath}[1]{\boldsymbol{#1}}

\begin{document}

\title{A Green's function approach to giant-dipole systems}

\author{Thomas Stielow, Stefan Scheel, Markus Kurz}

\address{Institut f\"ur Physik, Universit\"at Rostock, Albert-Einstein-Stra{\ss}e 23, D-18059 Rostock, Germany}

\ead{thomas.stielow@uni-rostock.de}

\begin{abstract}
In this work we perform a Green's function analysis of giant-dipole systems. First we derive the Green's functions of different magnetically field-dressed systems, in particular of electronically highly excited atomic species in crossed electric and magnetic fields, so-called giant-dipole states. We determine the dynamical polarizability of atomic giant-dipole states as well as the adiabatic potential energy surfaces of giant-dipole molecules in the framework of the Green's function approach. Furthermore, we perform an comparative analysis of the latter to and exact diagonalization scheme and show the general divergence behavior of the widely applied Fermi-pseudopotential approach. Finally, we derive the giant-dipole's regularized Green's function representation.
\end{abstract}

\section{Introduction}
Giant-dipole states are an exotic species of Rydberg atoms in crossed electric and magnetic fields. They have been first explored theoretically \cite{Baye1992,Dzyaloshinskii1992,Dippel1994,Schmelcher1998,Schmelcher2001} and experimentally \cite{Fauth1987,Raithel1993} firstly in the early 1990s. The giant-dipole states are of decentered character and possess, in contrast to the usual Rydberg states, a huge electric dipole moment. 
It has been shown that the total potential of the electronic motion possesses a gauge invariant term which leads to an outer potential well \cite{Schmelcher1993a}. This potential well supports weakly bound states possessing a large spatial separation of the Rydberg electron and the ionic core. The mathematical origin of this effect is the non-separability of the center of mass and electronic motions in the presence of the external fields \cite{Avron1978,Herold1981,Johnson1983,Ackermann1997,Shertzer1998}. More precisely, translation symmetry and conservation of the total momentum in field-free space is replaced by the conservation of a novel quantity, the so-called total pseudomomentum. Recently, the existence of giant-dipole Wannier excitons in crossed electric and magnetic fields have also been predicted inside a $\textrm{Cu}_2\textrm{O}$ semiconductor environment \cite{Kurz2017}. Furthermore, giant-dipole molecular states consisting of a highly excited atom in crossed fields and a neutral ground state atom have been shown to exist and the corresponding electronic configurations as well as their rovibrational properties have been analyzed in detail \cite{Kurz12}.

In the last years, emphasis has been laid on diatomic ultra-long range molecules consisting of a Rydberg atom whose electron binds a neutral ground state atom as a result low-energy scattering off the ground state species. These particular molecules were predicted and experimentally verified to exist in ultracold traps \cite{Greene2001,Hamilton2002,Stanojevic2006,Bendkowsky2009}. Since then a number of observations of Rydberg molecules  with a variety of electronic and rovibrational structures were made \cite{Tallant2012,Krupp2014,Gonzalez2014,Anderson2014,Booth2015,DeSalvo2015,Sassmannshausen2015,Sassmannshausen2016,Niederpruem2016,Niederpruem2016b,Camargo2016,Gonzalez2016}. The standard approach to describe these ultra-long range molecules is the Fermi-pseudopotential approach \cite{Fermi1934}. However, an alternative approach based on a Green's function method has been established as well \cite{Fabrikant2002,Greene2006,Bendkowsky2010,Fey2015,Fey2016}. In a recent work, C.\ Fey and coworkers pointed out the interconnections and limitations of both approaches \cite{Fey2015}. In particular, it was shown that the standard Fermi-pseudopotential approach modeled by a bare delta scattering potential leads to non-converging results. These difficulties can be overcome by considering a regularized $\delta$-function potential which is equivalent to the application of a regularized Green's function approach.

In this work, we apply the analysis performed in Ref.\ \cite{Fey2015} to the giant-dipole systems in general. After a short introduction, we present in Sec.\ 2 some general properties of the Green's function for the case of decoupled subsystems. In Sec.\ 3 we derive the Green's function for magnetically field-dressed systems including the giant-dipole system. The result of this analysis is then used in Sec.\ 4 to calculate a spectroscopic response function, namely the dynamical polarizability of giant-dipole systems. In Sec.\ 5 we analyze the adiabatic potential energy surfaces (PES) where we show a divergent behavior whenever a bare Fermi-pseudopotential approach is used to model the electron-perturber interaction. Using the result for the giant-dipole Green's function, we verify this divergent behavior for the giant-dipole system. In Sec.\ 6 we use the giant-dipole Green's function approach to calculate the PES in different levels of approximations. Finally, we present the result for the regularized giant-dipole Green's function.
\section{Green's function of decoupled systems \label{gd_intro}}
We first consider the Green's function of a system $H$ consisting of an arbitrary number $N$ of decoupled subsystems $h_i$. The related Green's function $G(\bfmath{r}, \bfmath{r}';E)$ of the total system $H$ in energy space is given by
\begin{eqnarray}
(E-H)G=\bfmath{1}\ \Rightarrow \ (E-H(\bfmath{r}))G(\bfmath{r}, \bfmath{r}';E)=\delta(\bfmath{r}-\bfmath{r}')\equiv \prod\limits^{N}_{i=1}\delta(\bfmath{r}_i-\bfmath{r}^{'}_{i}),\label{GFtotal}
\end{eqnarray}
where $H(\bfmath{r})=\sum_{i}h_i(\bfmath{r}_i),\ \bfmath{r}= (\bfmath{r}_1,...,\bfmath{r}_N)$. Using  the eigenfunctions $\phi_i$ of the subsystems $h_i$
\begin{eqnarray}
h_i \phi^{(i)}_{k}({\bf r}_i)=\varepsilon^{(i)}_{k}\phi^{(i)}_{k}({\bf r}_i),\ \ \ i=2,...,N,\label{eigen}
\end{eqnarray}
we can expand the Green's function $G$ of the total system as
\begin{eqnarray}
G(\bfmath{r},\bfmath{r}';E)=\sum_{k_2...k_N}G_{1}(\bfmath{r}_1,\bfmath{r}_1';E-\sum^{N}_{i=2}\varepsilon^{(i)}_{k_i}) \prod^{N}_{n=2} \phi^{(n)*}_{k_n}(\bfmath{r}^{\prime}_n)\phi^{(n)}_{k_n}(\bfmath{r}_{n}) \label{GFtotalexp}
\end{eqnarray}
whereby $G_1$ is the Green's function related to the Hamiltonian $h_1$, i.e.\
\begin{eqnarray}
(E-h_1)G_{1}(\bfmath{r}_1,\bfmath{r}_1';E)=\delta(\bfmath{r}_1-\bfmath{r}^{'}_{1}).\label{delta}
\end{eqnarray}
Using the completeness relation of the $\phi^{(i)}_k$-functions
\begin{eqnarray}
\delta(\bfmath{r}_i-\bfmath{r}^{'}_{i})=\sum_{k}\phi^{(i)*}_{k}(\bfmath{r}_i) \phi^{(i)}_{k}(\bfmath{r}^{'}_i)
\end{eqnarray}
one can directly verify that Eq.\ (\ref{GFtotalexp}) solves Eq.\ (\ref{GFtotal}) (see Appendix A). Furthermore, from Eq.\ (\ref{GFtotal}) one obtains the known spectral representation of an arbitrary Green's function $G$
\begin{eqnarray}
G(\bfmath{r},\bfmath{r}';E)=\sum_i \frac{\psi_i(\bfmath{r})\psi^{*}_{i}(\bfmath{r}^\prime)}{E-\varepsilon_i}, \label{GFeigenexp}
\end{eqnarray}
where $\psi_i$, $\varepsilon_i$ denote the eigenfunctions and eigenvalues of $H$. 

Obviously, the expansion Eq.\ (\ref{GFtotalexp}) is not unique as any subsystem $h_i$ can be used to construct the Green's function of the total system. Furthermore, it is easy to verify that in case of an unitary transformation $\mathcal{U}$ of the Hamilton $H$, the corresponding Green's function transforms analogously, i.e.\ $\tilde{G}(\bfmath{r},\bfmath{r}';E)=\mathcal{U}G(\bfmath{r},\bfmath{r}';E)\mathcal{U}^{-1}$.
\section{Green's functions of magnetically field-dressed systems}
We first derive the Green's function of various magnetically field-dressed systems including the considered giant-dipole system.
\subsection{Harmonically trapped systems}
As a first example we consider a particle with charge $q$ of mass $\mu$ in a homogeneous magnetic field \cite{atomicunits} along the $z$-direction ($\bfmath{ B}=B\bfmath{e}_z$), confined to an additional harmonic potential in $z$-direction
\begin{equation}
H=H_{xy}+H_z,\ \ H_{xy}=\frac{1}{2 \mu }\left(\bfmath{p}^{2}_{x}+\bfmath{p}^{2}_{y} \right)-\frac{qB}{2 \mu}L_{z}+\frac{q^2B^2}{8 \mu^2}(x^2+y^2),\ \ H_z=\frac{p^{2}_{z}}{2 \mu}+\frac{\mu}{2}\omega^{2}_{z}z^2.\label{Hxy}
\end{equation}
In order to determine the Green's function related to this system we use the Green's function $G_{xy}$ of a magnetically field dressed charge in two dimensions. This is given by \cite{Bychov1960,Ueta1992}
\begin{eqnarray}
G_{xy}(x,y,x',y';E)=e^{i(x'y-xy')\mu \omega_c/2}G_{0}(\eta;E),\ \ \ \omega_c=\frac{qB}{\mu},
\end{eqnarray}
where we have introduced $\eta=((x-x')^2+(y-y')^2)\mu \omega_c/2$ and
\begin{eqnarray}
G_{0}(\eta;E)=e^{-\eta /2} \frac{\mu \omega_c}{2\pi}\sum^{\infty}_{n=0}\frac{L_{n}(\eta)}{E-\omega_c (n+\frac{1}{2})}. 
\end{eqnarray}
Here the expression $L_{n}(x)$ denotes the Laguerre polynomials \cite{Abramowitz1972}. Using both the eigenenergies $\varepsilon_n=\omega_z(n+1/2)$ and eigenfunction $\phi_n(z)=(\mu \omega_z/\pi)^{-1/4}(2^n n!)^{-1/2}$ $H_{n}(\sqrt{\mu \omega_z}z)\exp(-\mu \omega_zz^2/2)$ of the harmonic oscillator, we easily obtain the Green's function $G_{xy}$ of this system as
\begin{eqnarray}
G_{xy}(\bfmath{r},\bfmath{r}';E)=e^{i(x'y-xy')\mu \omega_c/2-\eta /2}\frac{\mu \omega_c}{2\pi}\sum^{\infty}_{n,m=0}\frac{ \phi_m(z)\phi_{m}(z')}{E-\omega_z(m+1/2)-\omega_c(n+\frac{1}{2})}L_{n}(\eta).
\end{eqnarray}
Next, we consider a situation with an additional anisotropic harmonic confinement both in $x$- and $y$-directions, whereby the harmonic confinement remains untouched. For such a configuration the Hamilton $H_{xy}$ from Eq.\ (\ref{Hxy}) is rewritten and given by
\begin{eqnarray}
H_{xy}=\frac{1}{2 \mu }\left( \bfmath{p}^{2}_{x}+\bfmath{p}^{2}_{y} \right)-\frac{qB}{2 \mu}L_{z}+\frac{q^2 B^2}{8 \mu^2}(x^2+y^2)+\frac{\mu}{2}\omega^{2}_{x}x^2+\frac{\mu}{2}\omega^{2}_{y}y^2,\ \ \omega_x \not= \omega_y.\label{gdxy}
\end{eqnarray}
In contrast to Eq.\ (\ref{Hxy}) this Hamiltonian has lost its azimuthal symmetry. However, in Ref.\ \cite{Dippel1994} it has been shown that such a Hamiltonian can be diagonalized via an unitary transformation 
\begin{eqnarray}
U=\exp(i \alpha xy)\exp(i \beta p_x p_y),\ \alpha ,\beta \in \mathbb{R}
\end{eqnarray}
which transforms $H_{xy}$ into the sum of two decoupled harmonic oscillators with rescaled masses $M_{1,2}$ and frequencies $\omega_{1,2}$ with
\begin{eqnarray}
M_{1,2}&=&\frac{\sqrt{(\omega^{2}_{x}+\omega^{2}_{y}+\omega^{2}_{c})^2-4\omega^{2}_{x}\omega^{2}_{y}}}{\textrm{sgn}(\omega^{2}_{x}-\omega^{2}_{y})(\omega^{2}_{x}-\omega^{2}_{y} \pm \omega^{2}_{c})+\sqrt{(\omega^{2}_{x}+\omega^{2}_{y}+\omega^{2}_{c})^2-4\omega^{2}_{x}\omega^{2}_{y}}},\\
\omega_{1,2}&=&\frac{1}{\sqrt{2}}[\omega^{2}_{x}+\omega^{2}_{y}+\omega^{2}_{c} \pm \textrm{sgn}(\omega^{2}_{x}-\omega^{2}_{y})\sqrt{(\omega^{2}_{x}+\omega^{2}_{y}+\omega^{2}_{c})^2-4\omega^{2}_{x}\omega^{2}_{y}}]^{1/2}.
\end{eqnarray}
Thus, the  eigenenergies $\varepsilon_{n_1 n_2}$ and eigenfunctions $\psi_{n_1 n_2}(x,y)$ of Eq.\ (\ref{gdxy}) are given by
\begin{eqnarray*}
\varepsilon_{n_1 n_2}=(n_1+1/2)\omega_1+(n_2+1/2)\omega_2,\ \ \psi_{n_1 n_2}(x,y)=U(\phi_{n1}(x)\phi_{n2}(y)),\ n_{i} \in \mathbb{N}_{0}.\label{harmonic_oscillator_prop}
\end{eqnarray*}
The functions $\phi_{n_1}(x),\phi_{n_2}(y)$ denote the eigenfunctions of one-dimensional harmonic oscillator with masses $M_i$ and frequencies $\omega_i$. The specific form of the eigenfunction $\psi_{n_1n_2}(x,y)$ depends on $U$ and has been derived explicitly in Ref.\ \cite{Dippel1994}. These eigenenergies and eigenfunctions can be used to expand the Green's function related to the Hamiltonian $H=H_{xy}+H_z$ from Eq.\ (\ref{Hxy}) by using the Green's function of the one-dimensional harmonic oscillator which is given by \cite{Bakhrakh1970}
\begin{eqnarray}
G^{\rm (1d)}_\text{har}(z, z'; E) = - \sqrt{\frac{\mu}{\pi \omega_z}} \, \Gamma\left(\frac 1 2 - \frac E {\omega_z}\right) \, D_{E / \omega_z - 1 / 2}(-\sqrt{2 \mu E} \, z_<) \, D_{E / \omega_z - 1 / 2}(\sqrt{2 \mu E} \, z_>).
\end{eqnarray}
Here $D_{E}$ denotes a parabolic cylinder function \cite{Abramowitz1972} and $z_{<},\ z_{>}$ are the lesser and greater of the spatial variables $z$ and $z'$, $z_{<} \equiv \textrm{min}(z,z'),\ \ z_{>} \equiv \textrm{max}(z,z')$. Using Eq.\ (\ref{GFtotalexp}) we easily obtain
\begin{eqnarray}
\mathcal{G}_{\rm gd}(\bfmath{r},\bfmath{r}';E)=\sum_{n_1 n_2}G^{\rm (1d)}_{\textrm{har}}(z,z',E-\varepsilon_{n_1n_2})\psi_{n_1n_2}(x,y)\psi^{*}_{n_1n_2}(x',y').\label{gdgf} 
\end{eqnarray}
The poles of the Green's function (\ref{gdgf}), which are related to the poles of the Gamma function $\Gamma$, determine the eigenenergies of the total system $H$. As $\Gamma(x)$ possesses poles for $x=-n,\ n \in \mathbb{N}_0$, the eigenenergies of $H$ are given by $E_{n_1n_2n}=\varepsilon_{n_1n_2}+\omega_z(n+1/2)$. This is the same result as it is expected from the structure of the transformed Hamiltonian $\tilde{H}=UHU^{-1}$.
\subsection{Giant-dipole Green's function \label{gd_gf}}
The main focus this work are a special kind of atomic species, the so-called giant-dipole atoms which consists of a highly excited atom in crossed homogeneous electric and magnetic fields. In particular, we consider the case that a highly excited single valence electron orbits around a positively charged ionic core of coordinate $\bfmath{r}_c$ and mass $m_c$. If we approximate the ionic core as a charged point particle, we find the following Hamiltonian
\begin{eqnarray}
H=\frac{(\bfmath{p}+\bfmath{A}(\bfmath{r}_e))^2}{2m_e} + \frac{(\bfmath{p}-\bfmath{A}(\bfmath{r}_c))^2}{2m_c}+\bfmath{E}\cdot(\bfmath{r}_e-\bfmath{r}_c)-\frac{1}{|\bfmath{r}_e-\bfmath{r}_c|},\ \ \bfmath{A}(\bfmath{r})=\frac{1}{2}\bfmath{B}\times \bfmath{r},\ \ \bfmath{B} \perp \bfmath{E}.\label{laber}
\end{eqnarray}
This system has been studied in detail in Refs.~\cite{Dippel1994,Schmelcher1998,Schmelcher1993a}. In systems of crossed electric and magnetic fields the total momentum $\bfmath{P}$ is not a conserved quantity, but the so-called pseudomomentum $\bfmath{K}\equiv\bfmath{P}-\bfmath{B}\times \bfmath{r}/2$ \cite{Dippel1994}. For this reason, Eq.\ (\ref{laber}) can be transformed into a single particle representation where the corresponding Hamiltonian is given by
\begin{eqnarray}
H_\text{gd}=\frac{(\bfmath{p}-q\bfmath{A}(\bfmath{r}))^2}{2\mu}+V_{\rm gd}(\bfmath{r}),\ \ V_{\rm gd}(\bfmath{r})=\bfmath{E}\cdot \bfmath{r}-\frac{1}{r}+\frac{1}{2M}(\bfmath{K}+\bfmath{B}\times  \bfmath{r})^2
\end{eqnarray}
with $\bfmath{r}=\bfmath{r}_e-\bfmath{r}_c$, $M=m_e+m_c$, $\mu=m_em_c/M$ and $q=(m_e-m_c)/M$. Obviously, this is the problem of a particle of mass $\mu$ and charged $q$ in an external magnetic field and potential $V_{\rm gd}$. Using the transformation $\bfmath{K}\rightarrow \bfmath{K}-M(\bfmath{E}\times \bfmath{B})/B^2$, the electric field can be incorporated into the pseudomomentum. Hence we can set $\bfmath{E}=0$ in the following and discuss the properties of the potential $V_{\rm gd}$ via $\bfmath{K}$ only. Note that $K=|\bfmath{K}|=1\ \rm a.u.$ corresponds to an electric field strength of $2.8 \cdot 10^3\ \rm V/m$ \cite{Dippel1994}.

For a specifically chosen set of parameters, this potential exhibits an outer well containing many bound electronic states. As atomic states bound in this outer potential well exhibit huge permanent electric dipole moments of around $10^5\, \textrm{Debye}$, they have become known as giant-dipole states \cite{Dippel1994}. Expanding the potential $V_{\rm gd}$ around the minimum \textbf{$\bfmath{r}_{\rm min}$} of the outer well up to second order
\begin{eqnarray}
V_{\rm gd}(\bfmath{r})=\frac{\mu}{2}\left( \omega^{2}_{x}(x-x_{\rm min})^2 + \omega^{2}_{y}(y-y_{\rm min})^2 + \omega^{2}_{z}(z-z_{\rm min})^2 \right) 
\end{eqnarray}
and performing a corresponding spatial transformation $\bfmath{r}\rightarrow \bfmath{r}+\bfmath{r}_{\rm min}$ \cite{Dippel1994} we arrive at the giant-dipole Hamiltonian which represents a charge $q$ in a homogeneous magnetic field $\bfmath{B}$ and a three-dimensional harmonic potential
\begin{eqnarray}
H_{\rm gd}= \frac{(\bfmath{p}-q\bfmath{A}(\bfmath{r}))^2}{2\mu}+\frac{\mu}{2}\left( \omega^{2}_{x}x^2 + \omega^{2}_{y}y^2 + \omega^{2}_{z}z^2 \right). 
\end{eqnarray}
The frequencies $\omega_{i}\equiv\omega_i(\bfmath{K},\bfmath{B})$ depend on the applied fields and characterize the anisotropy of the outer potential well. As $m_c \gg m_e$ one has $q \approx -1$ and $\mu \approx m_e$, which means that the giant-dipole system can be effectively described by an electron in a homogeneous magnetic field and the external potential $V_{\rm gd}$.

Throughout this work we consider $B=2.35\, \text{T},\ K=1\, \rm a.u.$. For arbitrary field strengths the Green's function related to the giant-dipole Hamiltonian $H_{\rm{gd}}$ is given by $\mathcal{G}_{\rm gd}$ in Eq.\ (\ref{gdgf}). However, for the fields  considered in this work one can verify that the system possesses nearly cylindrical symmetry, i.e.\ $\omega_x \approx \omega_y \equiv \omega_{\rho}$. If not stated otherwise, we adopt this azimuthal approximation throughout this work. In this case, it is straightforward to verify that the corresponding giant-dipole Green's function $G_{\rm gd}$ reads as 
\begin{eqnarray}
G_\text{gd}(\bfmath{r},\bfmath{r}';E)=\frac{1}{2\pi}\sum^{\infty}_{n=0}\sum^{\infty}_{m=-\infty}G^{\rm (1d)}_{\text{har}}(z,z',E-\varepsilon_{nm})R_{nm}(\rho)R_{nm}(\rho')e^{i m(\phi-\phi')}\label{gd_azymuthal} 
\end{eqnarray}
with
\begin{eqnarray}
R_{nm}(\rho)&=& \sqrt{\frac{2 \mu \Omega (n+|m|)!}{n!(|m|!)^2}}\ee^{-\frac1 2 \mu \Omega\rho^2} \left(\sqrt{\mu \Omega } \rho\right)^{|m|} {}_1F_1(-n;|m|+1;\mu \Omega\rho^2)
\end{eqnarray}
and
\begin{eqnarray}
\varepsilon_{nm}&=&2\Omega \left( n+\frac{|m|+1}{2} \right) + \omega_c \frac m 2.
\end{eqnarray}
Here ${}_1F_1$ denotes the confluent hypergeometric function \cite{Abramowitz1972} of the first kind, while $\omega_c = -B /\mu$ is modified cyclotron frequency and $\Omega^2=\omega_c^2/4+\omega^{2}_{\rho}$. 

According to Sec.\ \ref{gd_intro} we can rewrite Eq.\ (\ref{gd_azymuthal}) by using the Green's function of the two-dimensional isotropic harmonic oscillator $G^{\rm (2d)}_{\rm har}$ is given by \cite{Bakhrakh1972}
\begin{eqnarray}
G^{(\rm 2d)}_{\rm har}(\rho,\rho',\phi-\phi^\prime;E)=-\frac{1}{2 \pi \rho \rho^\prime}\sum^{\infty}_{m=-\infty}\frac{\Gamma \left( \frac{|m|+1-E}{2} \right)}{\Gamma \left( \frac{|m|+1}{2} \right)}M_{E/2,|m|/2}(\rho^{2}_{<})W_{E/2,|m|/2}(\rho^{2}_{>})e^{im(\phi-\phi^{\prime})},
\end{eqnarray}
where $M_{a,b}$ and $W_{a,b}$ are Whittaker functions \cite{Abramowitz1972} and $\rho_{<(>)}$ are, respectively, the lesser and the greater of $\rho$ and $\rho^\prime$. This gives an alternative expression to Eq.\ (\ref{gd_azymuthal}), namely
\begin{eqnarray}
G_\text{gd}(\bfmath{r},\bfmath{r}';E)=\sum^{\infty}_{n_z=0}G^{(\rm 2d)}_{\rm har}(\rho,\rho',\phi-\phi^\prime;E-\omega_z(n_z+\frac{1}{2}))\phi_{n_z}(z)\phi_{n_z}(z').
\end{eqnarray}
\section{Dynamic polarizabilities of giant-dipole states}
As a first application of the giant-dipole Green's function, we consider the dynamic polarizability $\alpha(\omega)$ that connects the system's induced dipole moment $d_{\rm ind}$ as a response to an external time-dependent driving field with frequency $\omega$, i.e.\ $\bfmath{d}_{\rm ind} = \underline{\bfmath{\alpha}}(\omega) \bfmath{E}(\omega)$. As the dynamic polarizability tensor $\underline{\bfmath{\alpha}}(\omega)$ is represented by a symmetric $3 \times 3$ matrix, the induced dipole moment and the corresponding energy shift is determined by three independent quantities according to the principal axis theorem. In case of the giant-dipole system the dynamic polarizability depends on the entries $\alpha_{ii},\ i=x,y,z$, which are the polarizabilities parallel or perpendicular to the magnetic field direction. In particular, it can be shown that $\alpha^{(n)}_{ii}$ is related to the system's Green's function via the relation \cite{Davydkin1971}  
\begin{eqnarray}
\alpha^{(nmn_z)}_{ii}(\omega)&=&-\sum_{\sigma = \pm 1}\langle \Psi_{nmn_z} |x_iG_{\rm gd}(\omega_{nmn_z} \pm \sigma \omega)x_i|\Psi_{nmn_z}\rangle \nonumber\\
&=&- \sum_{\sigma = \pm 1}\int {\bf d} \bfmath{r} \, {\bf d} {\bfmath{r}}^\prime \, \Psi^{*}_{nmn_z}(\bfmath{r}) x_i G_{\rm gd} \left( \bfmath{r}, \bfmath{r}^\prime; \omega_{nmn_z} + \sigma \omega \right) {x_i}^\prime \Psi_{nmn_z}(\bfmath{ r}^\prime).
\end{eqnarray}
where $\Psi_{nmn_z}$ and $\omega_{nmn_z}$ denote the giant-dipole eigenfunctions and eigenenergies, respectively. Expressing the confluent hypergeometric functions as generalized Laguerre polynomials \cite{Abramowitz1972} and using three term recursion formulas we find the polarizabilities parallel and perpendicular to $\bfmath{B}$ 
\begin{equation}
\alpha_n^\parallel(\omega) = \frac 1 \mu \frac{1}{\omega_z^2 - \omega^2},\ \ 
\alpha_n^\perp(\omega) = \frac{1}{2 \mu \Omega} \left(\frac{\Omega + \omega_c/2}{(\Omega + \frac 1 2 \omega_c)^2 - \omega^2} + \frac{\Omega - \omega_c/2}{(\Omega - \frac 1 2 \omega_c)^2 - \omega^2}\right) \, \text .\label{pol_dyn}
\end{equation}
For finite atomic lifetimes $\tau$, i.e.\ a decay rate $\Gamma=1/\tau$ for the giant-dipole states, one has to replace $\omega^2 \rightarrow \omega^2+i \omega \Gamma$. The imaginary part of the complex polarizability provides information about the dispersion properties. However, in this analysis we have taken the limit $\Gamma \rightarrow 0$.
\begin{figure}
\centering
\begin{minipage}[t]{0.49\textwidth} 
\includegraphics[width=\textwidth]{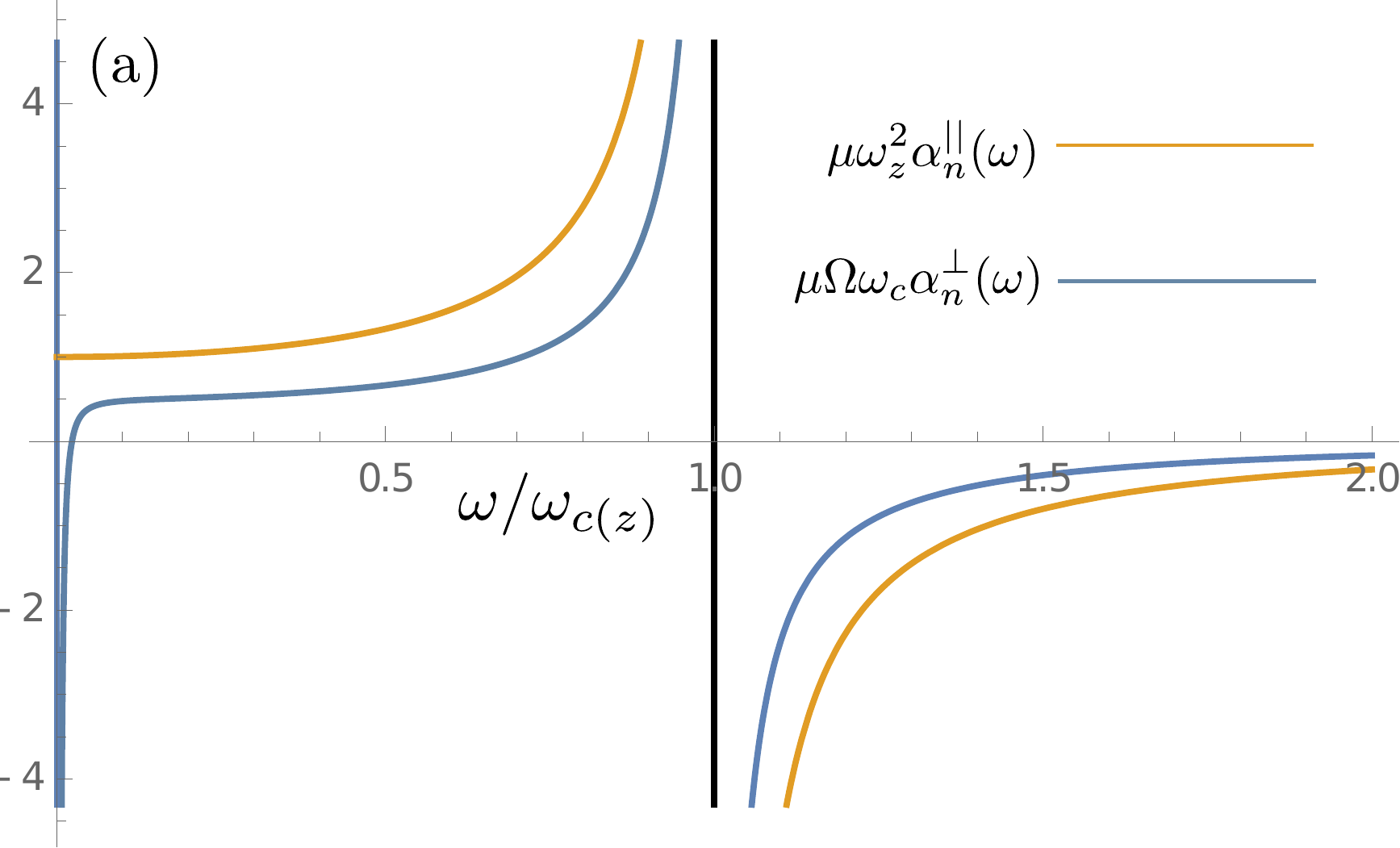}  
\end{minipage}
\begin{minipage}[t]{0.49\textwidth} 
\includegraphics[width=\textwidth]{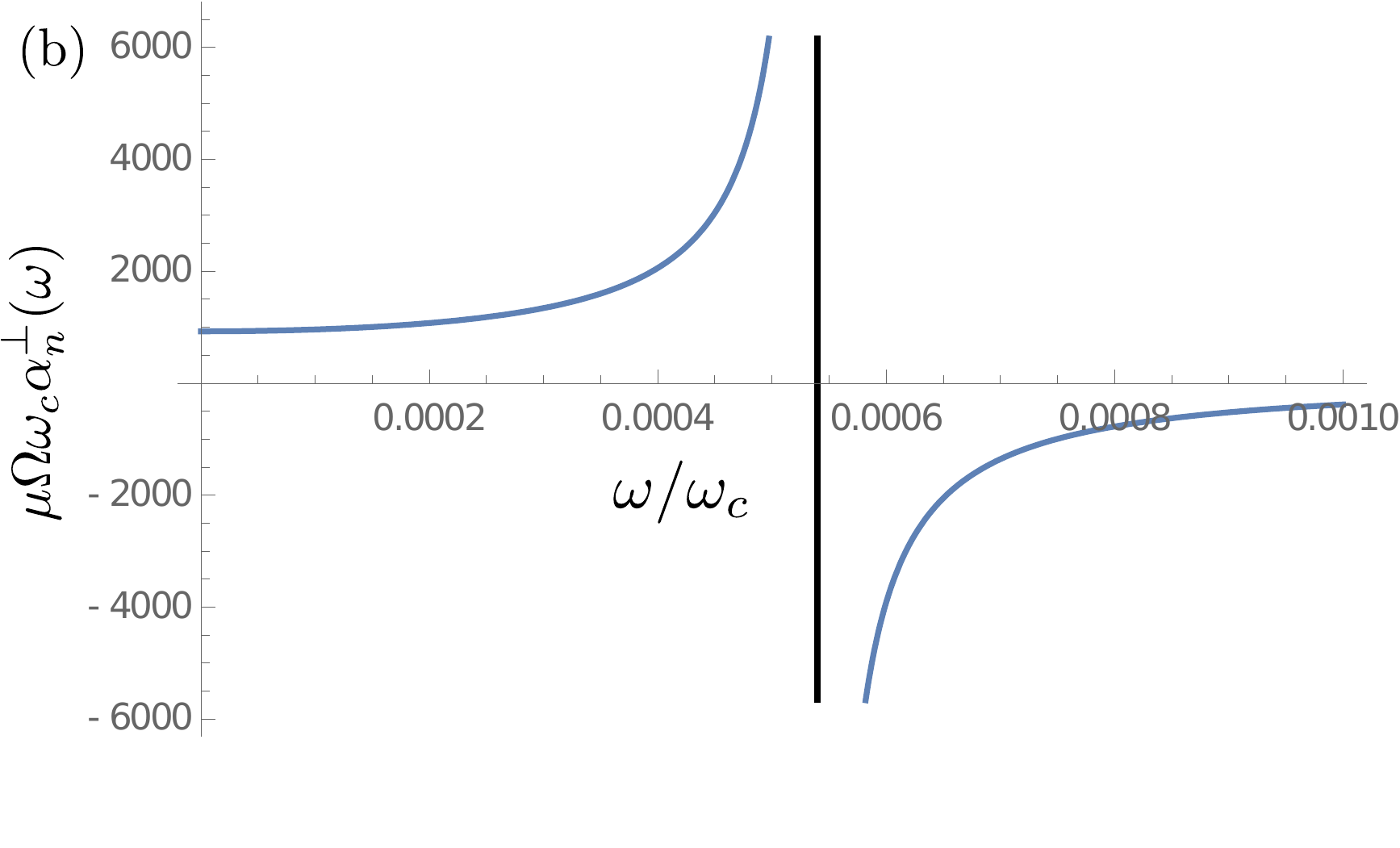} 
\end{minipage}
\caption{(a) Frequency dependent scaled dynamical polarizabilities $\mu  \omega^{2}_{z}\alpha^{||}_{n}(\omega)$ and $\mu \Omega \omega_c \alpha^{\perp}_{n}(\omega)$ as functions of $\omega/\omega_z$ and $\omega/\omega_c$, respectively. The black line near $\omega/\omega_{c(z)}=1$ indicates the divergence of the polarizabilities. (b) Magnified plot for the interval $\omega/\omega_c \in [0,10^{-3}]$ for $\mu \Omega \omega_c \alpha^{\perp}_{n}(\omega)$. The divergence around $\omega/\omega_c=5 \cdot 10^{-4}$ is clearly visible.}\label{dyn_pol}
\end{figure}

The polarizability parallel to the magnetic field is described by a Drude-Lorentz model, where a single valence electron is considered to be confined in a one-dimensional harmonic potential. In the limiting case of $\omega_c \to 0$, the same result is obtained for the polarizability in the perpendicular $\bfmath{B}$-field direction. The polarizabilities do not depend on the specific state, a result which is a typical feature of harmonically confined particles and also obtained in the Drude-model. This result can be extended beyond the azimuthal approximation, i.e.\ for arbitrary electric and magnetic field strengths. The explicit results of this analysis is presented in Appendix B. In Fig.\ \ref{dyn_pol} we show the frequency dependence of the scaled dynamical polarizabilities $\mu \omega^{2}_{z}\alpha^{||}_{n}(\omega)$ and $\mu \Omega \omega_{c}\alpha^{\perp}_{n}(\omega)$, respectively. Due to their dependence on $\omega_z, \omega_c$ and $\Omega$ the polarizabilities are implicit functions of the applied magnetic field strength and the pseudomomentum $\bfmath{K}$. In case of the parallel to $\bfmath{B}$ polarizability we clearly see the resonances for $\omega/\omega_z=1$. For the polarizability perpendicular to $\bfmath{B}$ we find four resonances at $\omega_{1,2}/\omega_c=(\Omega/\omega_c \pm 1/2)$, with give $\omega_1 \approx 1.00054$ and $\omega_{2} \approx 5.4 \times 10^{-4}$, respectively.

Note that the quantity $\alpha(\omega=0)$ commonly denotes the system's static polarizability which relates the induced dipole moment to a static electric field. However, for giant-dipole atoms it is not reasonable to define such a quantity since an additional external electric field simply redefines the system to a new giant-dipole species. 
\section{The electronic problem of ultra-long range molecules - Green's function ansatz \label{gf_div}}
The general problem of ultra-long range molecules consists of atomic species of highly excited electronic character possessing a single excited valence electron with a very low kinetic energy. The first molecular species under consideration were based on alkali systems with only one valence electron in the outermost electronic orbital \cite{Greene2001,Bendkowsky2009,Sassmannshausen2015}. However, in recent years molecules consisting of divalent constituents such as strontium \cite{DeSalvo2015,Camargo2016} have been realized as well.

The standard ansatz for the electronic Hamiltonian in adiabatic approximation is given by 
\begin{eqnarray}
H_\text{el}=H_0+V_\text{en}(\bfmath{r},\bfmath{R})\ \ \text{with}\ \ H_{0}=T+V(\bfmath{ r})\label{elecham}. 
\end{eqnarray}
Here, the coordinates $\bfmath{r}=(x,y,z)$ and $\bfmath{R}=(x_n,y_n,z_n)$ denote the spatial positions of the electron and the neutral ground state atom (neutral pertuber for short), respectively. $H_0$ is the Hamiltonian describing the Rydberg electron in its ionic core potential $V(\bfmath{r})$ and $V_{\rm en}(\bfmath{r},\bfmath{R})$ is the interaction between the Rydberg electron and the neutral perturber. Finally, we model the interatomic potential for the low-energy scattering between the Rydberg electron and the neutral perturber by the Fermi-pseudopotential \cite{Fermi1934,Omont1977}
\begin{eqnarray}
V_{\rm{en}}(\bfmath{r,R})=2\pi A_s[k]\delta(\bfmath{r}-\bfmath{R}).\label{fermi} 
\end{eqnarray}
In Eq.\ (\ref{fermi}), the quantity $A_s[k] = -\tan ( \delta_0( k ))/ k$ denotes the energy-dependent triplet $s$-wave scattering length which is evaluated from the corresponding phase shifts $\delta_0(k)$ \cite{Fabrikant2002}. The kinetic energy $E_{\rm{kin}} = k^2/2$ of the valence electron at the collision point with the neutral perturber can be taken in a semiclassical approximation $k^2/2 =E_{\rm{el}}-V(\bfmath{r})$.

Over time, two approaches have been established to deduce the electronic eigenenergies $\varepsilon(\bfmath{R})$ of $H_{\rm el}$, which parametically depend on the perturber position and serve as an input for its dynamics. First, the electronic Hamiltonian $H_{\rm{el}}$ can be diagonalized using an arbitrary single particle basis $\{ | \phi_i \rangle \}_{i=1,...,N}$. Depending on the specific problem, quantum defect rubidium wave function \cite{Greene2001}, giant-dipole states \cite{Kurz12} and hybridized hydrogen states \cite{Eiles2016} have been applied. In an alternative approach, the Green's function $G_{0}(\bfmath{r},\bfmath{r}^\prime)$ of the unperturbed electronic problem $H_0$ satisfying 
\begin{eqnarray}
(E-H_0)G_{0}(\bfmath{r},\bfmath{r}';E)=\delta(\bfmath{r}-\bfmath{r}')\label{gf_formula}
\end{eqnarray}
is used to formulate an implicit problem for the system's energy $E$. In particular, in Ref.\ \cite{Fey2015} it has been shown that this approach leads to the following self-consistent integral equation for the perturbed wave function and energies $E$, respectively
\begin{eqnarray}
\psi({\bf r})=-\int d \bfmath{r}' G_{0}(\bfmath{r},\bfmath{r}';E)V_{\rm{en}}(\bfmath{r}',\bfmath{R})\psi(\bfmath{ r}')\ \Rightarrow \ 1 - 2\pi A_s[k]G_{0}(\bfmath{R},\bfmath{R};E)=0.\label{wavefunc}
\end{eqnarray}
In previous works the Green's function approach has been applied to systems where the highly excited Rydberg electron is attached to an ionic core potential $V(r)=-1/r+V_{\rm{qd}}(r)$, where $V_{\rm{qd}}(r)$ is a short ranged quantum defect potential which takes into account deviations of the Coulomb potential due to finite size effects of the core of non-hydrogenic atomic species. It is known that for such systems the Green's function can be written as $G_{0}(\bfmath{r},\bfmath{r}';E)=G_{\rm{C}}(\bfmath{r},\bfmath{r}';E)+G_{\rm{qd}}(\bfmath{r},\bfmath{r}';E)$ whereby $G_{\rm{C}}$ denotes the Coulomb Green's function \cite{Fabrikant2002}. This Green's function diverges for $\bfmath{r} \rightarrow \bfmath{r}'$, in particular $G_{\rm{C}}(\bfmath{r},\bfmath{r}';E)\rightarrow -1/2\pi|\bfmath{r}-\bfmath{r}'|$ \cite{Fabrikant2002}. However, it can be shown that this kind of divergent behavior is not only restricted to Rydberg systems interacting via a Coulomb-like potential, but to all potentials $V(\bfmath{r})$ that are less divergent than a bare Coulomb potential.

To verify this prediction we introduce the relative position vector $\boldsymbol{\xi}=\bfmath{r}-\bfmath{r}'$, which gives $\Delta_{\bfmath{r}} \rightarrow \Delta_{\bfmath{\xi}}\equiv\partial^{2}_{\xi_1}+\partial^{2}_{\xi_2}+\partial^{2}_{\xi_3}$, $V(\bfmath{r})=V(\bfmath{\xi};\bfmath{r}^\prime)$ and $G_{0}(\bfmath{r},\bfmath{r}';E)=G_{0}(\bfmath{\xi},\bfmath{r}';E)$. For $|\boldsymbol{\xi}|=\xi \rightarrow 0$ we assume that $\xi^2 V \rightarrow 0$. In this case the kinetic energy term is dominant compared to the interaction potential $V$ in the Hamiltonian $H_0$ and $E$. From Eq.~(\ref{gf_formula}) we obtain
\begin{eqnarray}
\left( E-\frac{\bfmath{p}^2}{2}-V(\boldsymbol{\xi};\bfmath{r}^\prime) \right)G_{0}(\boldsymbol{\xi},\bfmath{r}^\prime;E)=\delta(\boldsymbol{\xi})\ \underset{\xi \rightarrow 0}{\Rightarrow} \ \frac{1}{2}\Delta_{\bfmath{\xi}} G_{0}(\boldsymbol{\xi},\bfmath{r}^\prime;E)=\delta(\boldsymbol{\xi}).\label{proof_1}
\end{eqnarray}
This problem is related to the Green's function of a free particle. $G_0$ is now a function of $\xi$ only, i.e. $G_{0}(\xi)$, and closely related to the Green's function of the three-dimensional Laplace operator which is given by $-1/(4\pi\xi)$. Thus, from Eq.\ (\ref{proof_1}) we easily derive 
\begin{eqnarray}
G_{0}(\xi) \approx -\frac{1}{2 \pi \xi} = -\frac{1}{2 \pi |\bfmath{r}-\bfmath{r}^\prime|}.\label{gf_limit_behavior}
\end{eqnarray}
Obviously, this is the functional behavior known from the Coulomb problem \cite{Fabrikant2002} as it has been predicted. We note that the $1/\xi$ dependence is not a direct result of the electron potential $V(\bfmath{r})$, but stems from the kinetic energy term. As the Green's function approach is equivalent to the exact diagonalization of the electronic Hamiltonian $H_0$ in Eq.\ (\ref{elecham}), Eqs.\ (\ref{wavefunc}) and (\ref{gf_limit_behavior}) indicate a non-convergent behavior in the case a Fermi-pseudopotential is used to model the electron-perturber interaction for any kind of adiabatic electron system.

For this reason, in order to study such systems, alternative approaches such as an exact diagonalization scheme using a regularized Fermi-pseudopotential as well as Green's function methods have been considered \cite{Fey2015,Fabrikant2002}. In the case that the regularized Green's function method is employed, one has to determine the energy $E$ from the self-consistent equation
\begin{eqnarray}
1 - 2\pi A_s[k]G^{\rm (reg)}_{0}(\bfmath{R},\bfmath{R};E)=0.\label{Greg_E}
\end{eqnarray}
The function $G^{\rm (reg)}_{0}$ denotes the systems regularized Green's function. Throughout the literature a number of regularization procedures have been discussed \cite{Fey2015, Fabrikant2002}. Here we follow the procedure applied by Fabrikant {\it et al.} \cite{Fabrikant2002} and define
\begin{eqnarray}
G^{\rm (reg)}_{0}(\bfmath{r};\bfmath{r}^{\prime};E)\equiv G_{0}(\bfmath{r};\bfmath{r}^{\prime};E)+\frac{1}{2 \pi |\bfmath{r}-\bfmath{r}^\prime|}\ \ \textrm{with}\ \  G^{\rm (reg)}_{0}(\bfmath{R},\bfmath{R};E) \equiv \lim_{\bfmath{r} \rightarrow \bfmath{R}} G^{\rm (reg)}_{0}(\bfmath{r};\bfmath{R};E),
\end{eqnarray}
which means that one simply subtracts the divergent term from $G_0$. This definition slightly differs from the definition in Ref.\ \cite{Fey2015}, but gives identical results in the case of $\bfmath{r}=\bfmath{r}^\prime= \bfmath{R}$ which is required for the determination of $E$ via Eq.\ (\ref{Greg_E}) (see Appendix C).
\section{Ultra-long range giant-dipole molecules}
In this section we consider a highly excited hydrogen atom which is prepared in a giant-dipole state interacting with a neutral ground state $^{87}$Rb atom (see Sec.\ \ref{gf_div}). For this particular system, the molecular Hamiltonian reads
\begin{eqnarray}
H=\frac{\bfmath{P}^2}{2m_n}+H_{\textrm{gd}}+V_{\textrm{gd,n}}(\bfmath{r},\bfmath{R}),\label{gd_ham}
\end{eqnarray}
where $V_{\textrm{gd,n}}$ denotes the interaction of the neutral perturber atom with the giant-dipole Rydberg electron. For deeply bound states in the outer potential well the electron possesses a low kinetic energy. For this reason, it is legitimate to model this interaction potential via a Fermi-type $s$-wave pseudopotential as it has been discussed in Sec.\ \ref{gf_div}. Analogous to previous works we consider the electron-perturber interaction to be purely determined by the triplet scattering channel \cite{Kurz12}. 
In order to solve the eigenvalue problem associated with Hamiltonian (\ref{gd_ham}) an adiabatic ansatz is employed for the electronic and perturber degrees of freedom, leading to the electronic Hamiltonian $H_{\textrm{el}}=H_{\textrm{gd}}+V_{\textrm{gd,n}}$ which parametically depends on the perturber position. In order to obtain the PES, an exact diagonalization scheme using the eigenstates of $H_{\textrm{gd}}$ has been adopted in a previous analysis \cite{Kurz12}. As the giant-dipole Hamiltonian $H_{\textrm{gd}}$ describes an magnetically-field dressed electron in an external three dimensional harmonic potential, its eigenfunctions are identical to those of Eq.\ (\ref{harmonic_oscillator_prop}). 

However, in this work we employ the Green's function approach as it was presented in Sec.~\ref{gd_intro} and 3, respectively. In this case, we have to determine the energies $E$ which are the roots of the equation 
\begin{eqnarray}
1-2\pi A_s[k]G_{\textrm{gd}}(\bfmath{R},\bfmath{R};E)=0, \label{eq_gf_E}
\end{eqnarray}
which has to be determined via a numerical root-finding routine. As discussed in Section \ref{gf_div} the Green's function $G_{\rm gd}$ of the giant-dipole system for arbitrary field strengths would be given by Eq.\ (\ref{gdgf}), i.e. $G_{\rm gd}=\mathcal{G}_{\rm gd}$.

As it has been discussed in Sec.~\ref{gf_div} the bare giant-dipole Green's functions \eqref{gd_azymuthal} possess a singularity for $\bfmath{r}=\bfmath{r}'$. This behavior can easily be verified for $\bfmath{r}'=0$:
\begin{eqnarray}
\lim\limits_{\bfmath{r}\rightarrow 0}G_{\rm gd}(\bfmath{r},0,E)\sim \lim\limits_{z\rightarrow 0} \sum^{\infty}_{n=0} G^{\rm (1d)}_{\rm har}(z,0,E-\varepsilon_{n0}) \sim -\sum^{\infty}_{n_z=0}\phi^{2}_{2n_z}(0)\sum^{\infty}_{n=0}\frac{1}{n}\rightarrow -\infty.
\end{eqnarray}
According to the analysis provided in Sec.~\ref{gf_div} this divergence is supposed to behave as $-1/(2\pi r)$. To analyze this in more detail, we introduce the truncated giant-dipole Green's function $G^{\rm (tr)}_{\rm gd}$ via its spectral representation, i.e. 
\begin{equation}
G^{\rm (tr)}_\text{gd}(\bfmath{r}, \bfmath{r}^\prime; E;\bfmath{N}) = \frac{1}{2\pi}\sum_{n=0}^N \sum_{m=-M_{\rm min} }^{M_{\rm max}} \sum_{n_z = 0}^{N_z}\frac{\phi_{n_z}(z)\phi_{n_z}(z') \, R_{nm}(\rho) R_{nm}(\rho') \, \ee^{i m \, (\phi - \phi')}}{E - \varepsilon_{nm} - \varepsilon_{n_z}}, \label{gf_trunc}
\end{equation}
with $\bfmath{N}=\{ M_{\rm min}, M_{\rm max}, N_z,N \}$. In case we restrict the summations up to a set of finite $\bfmath{N}$ we can analyze different levels of approximations of the exact function $G_{\rm gd}$. In particular, using the truncated Green's function Eq.\ (\ref{gf_trunc}) we can analyze the divergent behavior in more detail as for this approximation the divergence for $\bfmath{r}=\bfmath{r}^\prime$ is cut off.

\begin{figure}
\centering
\begin{minipage}[t]{0.455\textwidth} 
\includegraphics[width=\textwidth]{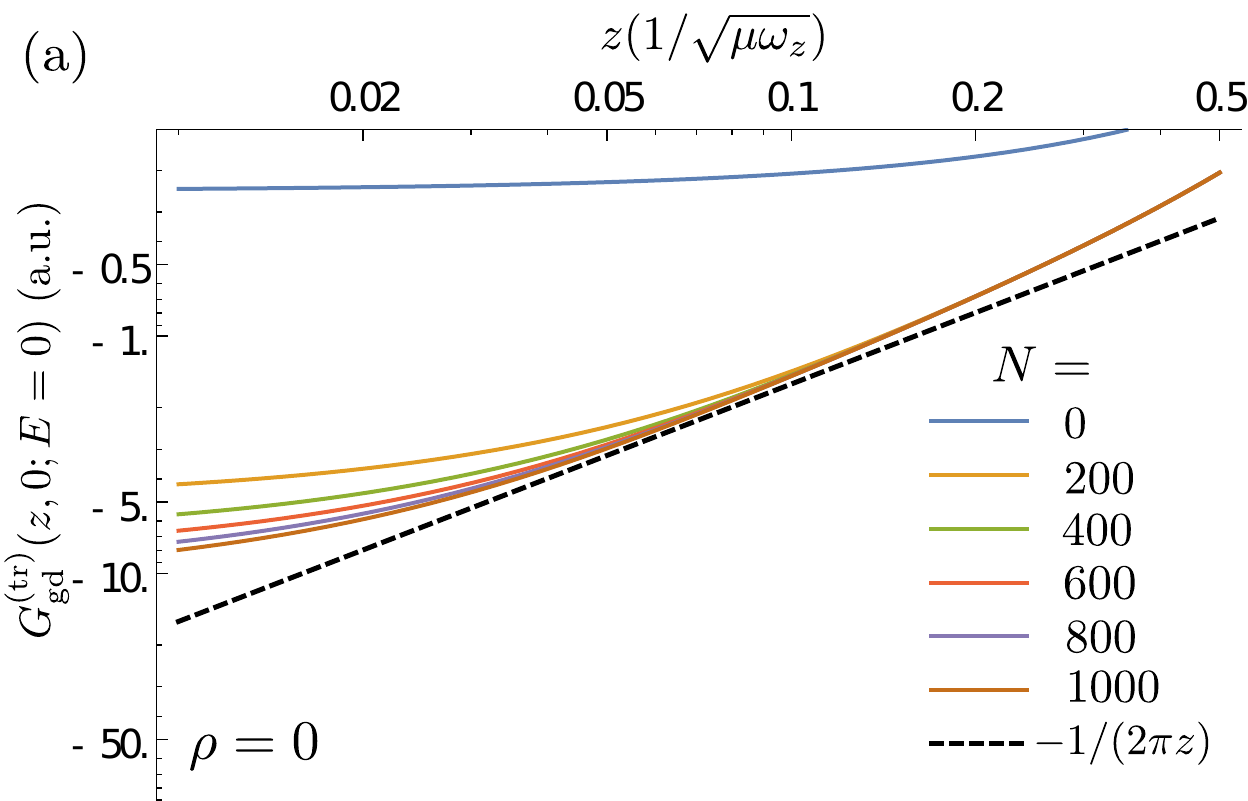}  
\end{minipage}
\begin{minipage}[t]{0.535\textwidth} 
\includegraphics[width=\textwidth]{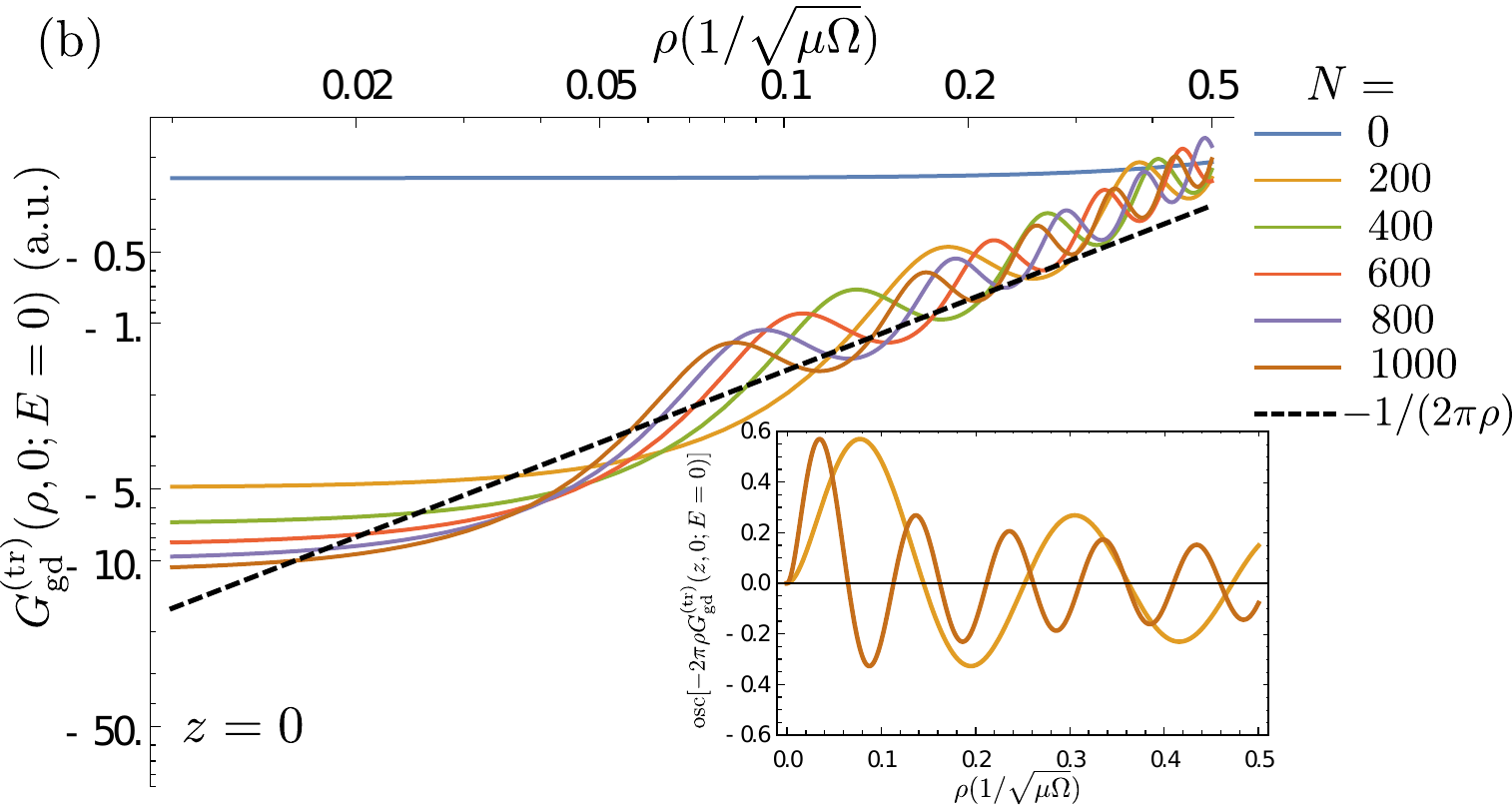} 
\end{minipage}
\caption{Double logarithmic plot for the asymptotic behavior of the truncated giant-dipole Green's function $G^{\rm (tr)}_{\rm gd}(\bfmath{r},0; E)$ given by Eq.\ (\ref{gf_trunc_2}) for $N=0,200,...,1000$ in the limit $\bfmath{r} \to 0$. Fig.\ (a) shows $G^{\rm (tr)}_{\rm gd}$ as a function of $z$ for $\rho=0$, where Fig.\ (b) presents the same function dependent on $\rho$ for $z=0$. In both figures the dashed black line indicates the limiting case $-1/(2\pi r)$. The inset in figure (b) shows the oscillations of $-2 \pi \rho G^{\rm (tr)}_{\rm gd}(\rho,0; E)$ for $N=200$ and $N=1000$, respectively.}
\label{fig:GreenDivergence3D}
\end{figure}
A detailed analysis of this issue is presented in Fig.\ \ref{fig:GreenDivergence3D}(a,b), where we present the functional behavior of the truncated Green's function for different levels of approximations. In particular, we have chosen the spatial coordinates to be $\bfmath{r}^\prime=0,\ \phi=0$ and take the limit $N_{z}\rightarrow \infty$. In this case the summation over $m$ is restricted to $m=0$ and we obtain
\begin{eqnarray}
G^{\rm (tr)}_\text{gd}(\rho,z,0; E)= \frac{1}{2 \pi}\sum^{N}_{n=0}G^{\rm (1d)}_{\rm har}(z,0;E-\varepsilon_{n0})\label{gf_trunc_2}R_{n0}(\rho)R_{n0}(0).
\end{eqnarray}
In Fig.\ \ref{fig:GreenDivergence3D}(a), we depict a double logarithmic plot for $G^{\rm (tr)}_{\rm gd}(0,z,0;E=0)$ for a number of different $N=0,200,...,1000$ as a function of $z$. In this representation the limiting curve $-1/(2 \pi z)$ is simply a straight line (black dashed curve) which makes it very convenient for comparison with the calculated data. Obviously, for increasing $N$ the Green's function approaches the divergent behavior as the numerical curves more and more approach the black dashed curve in the vicinity of $z \rightarrow 0$.

In Fig.\ \ref{fig:GreenDivergence3D}(b) we present the same analysis for $z=0$ following $\rho \rightarrow 0$, i.e.\ $G^{\rm (tr)}_{\rm gd}(\rho,0,0;E=0)$. Here, the situation is more complicated as the Green's function possess as strongly oscillating structure as a function of $\rho$. For increasing $N$ the numerical curves decrease approaching the $-1/2\pi \rho$ limit (black dashed curve) for $\rho \rightarrow 0$. However, the oscillations move inwards as well, making a precise analysis of the $N \rightarrow \infty$ behavior quite challenging. To get more insight we have analyzed the oscillatory behavior separately which is presented in the inset of Fig.\ \ref{fig:GreenDivergence3D}(b). Here, we show the oscillations of $N=200$ and $N=1000$ numerical curve. Obviously, for larger $N$ the frequency of the oscillating has increased while its amplitude decreases for sufficiently large $\rho$. In Appendix D we present a more systematic analysis where we show that the oscillatory behavior is well approximated by $J_{1}(2 \sqrt{N}\rho)$ where $J_{1}(x)$ denotes the $J_1$-Bessel function \cite{Abramowitz1972}. As $J_{1}(2 \sqrt{N}\rho) \rightarrow 0$ for  $N \rightarrow \infty$ we see that the oscillations vanish in the case of the full giant-dipole Green's function possessing a $-1/(2\pi r)$ behavior.
   
\begin{figure}
\centering
\begin{minipage}[t]{0.495\textwidth} 
\includegraphics[width=\textwidth]{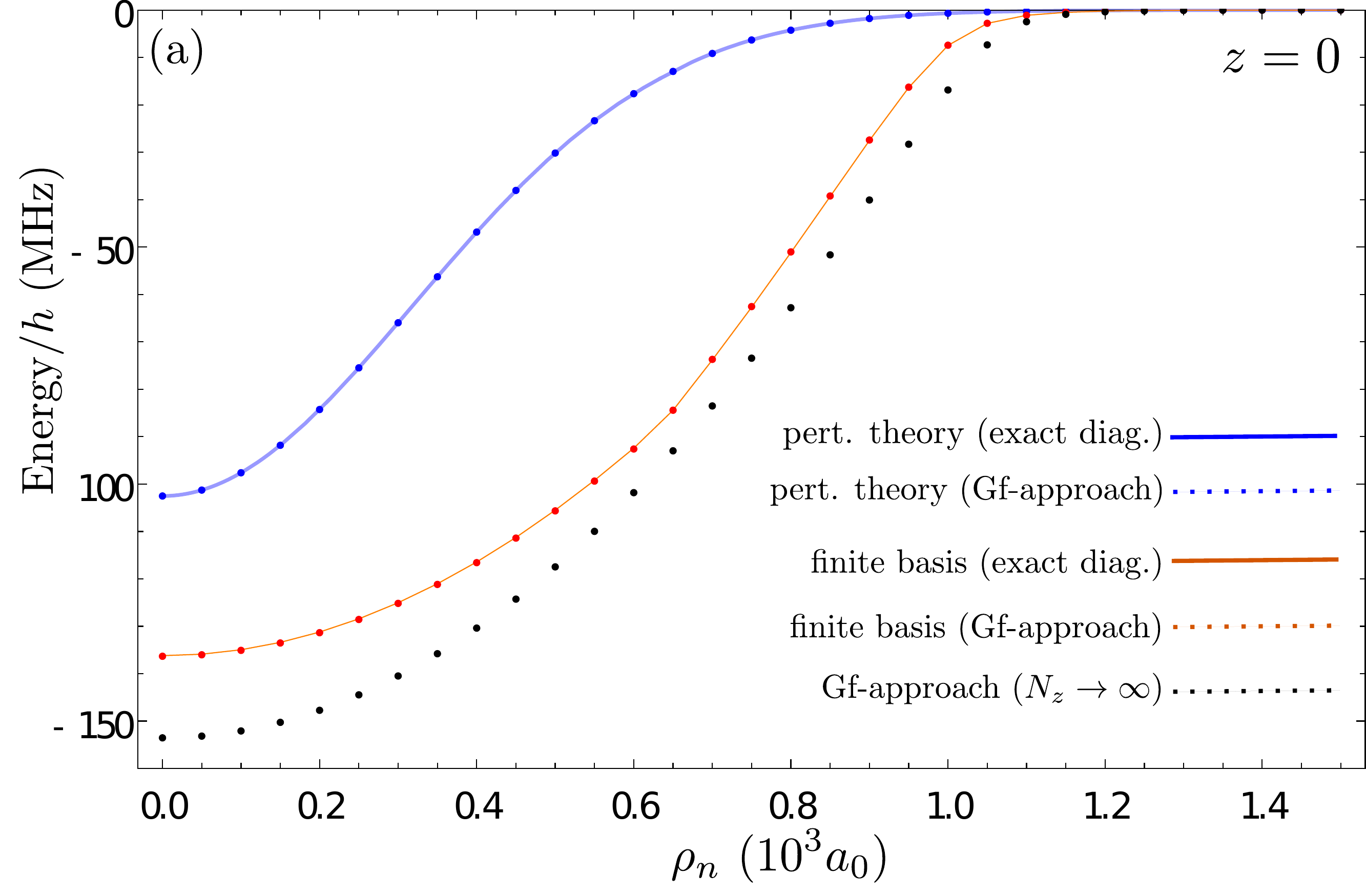}  
\end{minipage}
\begin{minipage}[t]{0.495\textwidth} 
\includegraphics[width=\textwidth]{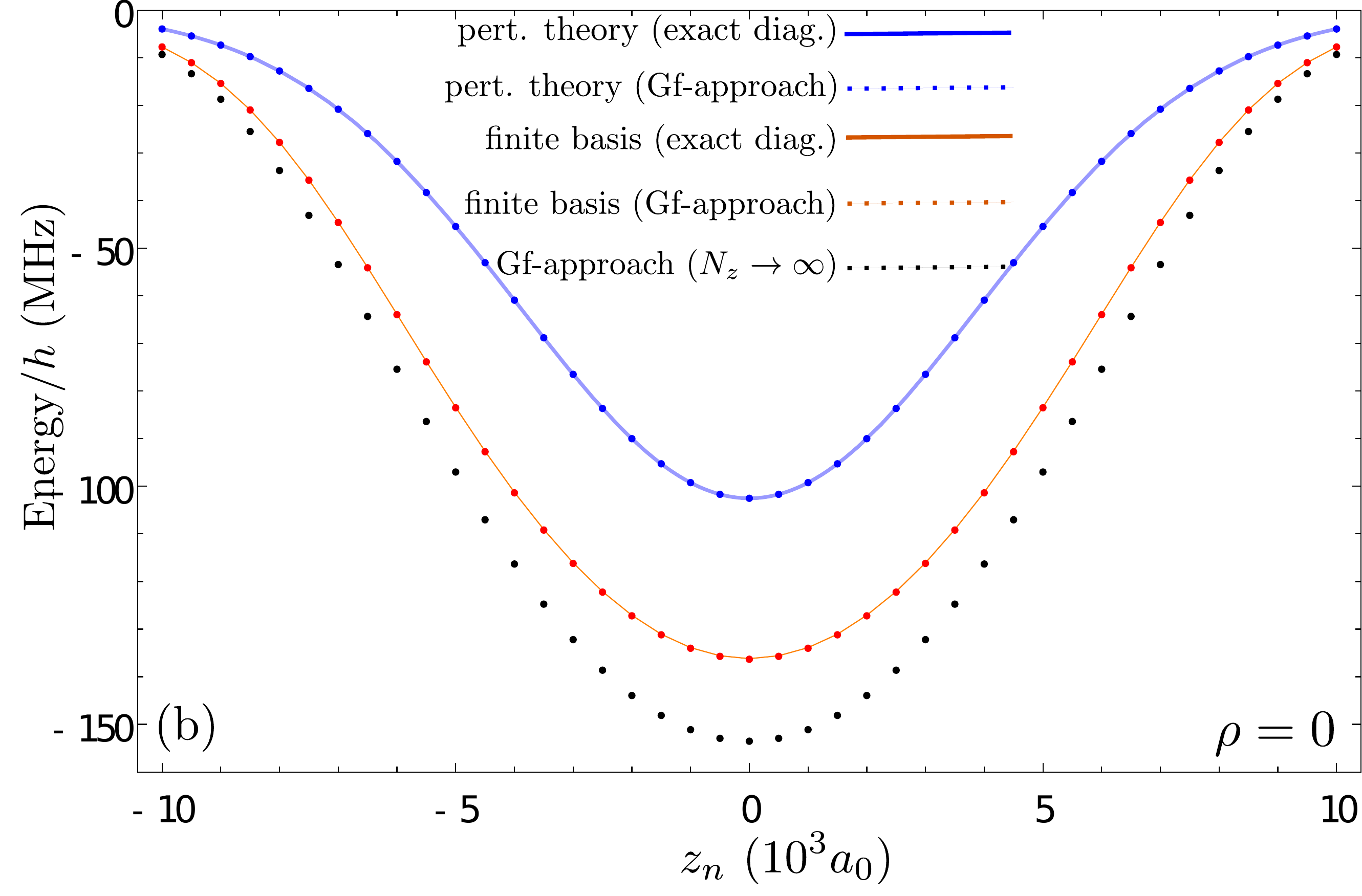} 
\end{minipage}
\caption{Ground state PES of a giant-dipole ultra-long range molecule with a rubidium perturber for $B = 2.35 \, \text T$, $K = 1.0$. Figure (a)((b)) show potential curves for $z_n=0$ ($\rho_n=0$). Both figures present results for different approaches: blue curve - first order perturbation theory, blue dots - perturbative Green's function, orange curve - exact diagonalization with $N_{z}=30$, $M_{\rm min} = 60$, $N=M_{\rm max}=0$, red dots - truncated Green's function approach Eq.\ (\ref{gf_trunc_2}) with the same set of functions, black dots - Green's function approach with $N_z \rightarrow \infty$ (Eq.~ (\ref{gf_trunc_3})).}\label{fig:GreenDiagCompareGD}
\end{figure}

Although the full Green's function possesses a divergent behavior, the truncated representation Eq.\ (\ref{gf_trunc}) can still be used to calculate approximate PES when it is inserted into Eq.\ (\ref{eq_gf_E}). As it has been discussed in \cite{Fey2015}, this is equivalent to the exact diagonalization approach where one uses a finite set of giant-dipole basis functions $\{ \Psi_{nmn_z}(\bfmath{r}) \}$ with $n=0,...,N$, $m=-M_{\rm min},...,M_{\rm max}$, $n_{z}=0,...,N_z$ to diagonalize the electronic Hamiltonian $H_{\rm el}$ given in Eq.~(\ref{elecham}). In the present case, we have calculated the ground state PES both via exact diagonalization and, in comparison, via the truncated Green's function approach using the same basis functions. The result of this analysis is presented in Figs.\ \ref{fig:GreenDiagCompareGD}(a,b), respectively. 

In particular, Fig.\ \ref{fig:GreenDiagCompareGD}(a) shows the ground state PES for $z_n=0$ as a function of $\rho_n$ for different levels of approximations. The blue solid line labeled as ``pert.\ theory (exact diag.)'' shows the pure first order perturbation theory result for the giant-dipole state $\Psi_{000}(\bfmath{r})$, i.e.\ 
\begin{eqnarray}
E(\rho_n,z_n)=\varepsilon_{00}+\frac{\omega_z}{2}+A_{s}[k]\phi^{2}_{0}(z_n)R^{2}_{00}(\rho_n). 
\end{eqnarray}
The same result is easily derived from Eq.\ (\ref{eq_gf_E}) for $M_{\rm min}=N_{z}=M_{\rm max}=N=0$, which is indicated by the blue dots and labeled as ``pert.\ theory (Gf-approach)''. In the same figure the solid red curve shows the exact diagonalization results  for a basis set of $M_{\rm min}=60,\ N_{z}=30$ and $M_{\rm max}=0,\ N=0$, the same parameters which had been employed for the truncated Green's function ansatz. The results of those calculation are indicated by red dots and labeled as ``finite basis (exact diag.)'', in comparison to the results of the exact diagonalization approach which is labeled as ``finite basis (Gf-approach)''. We see that both results are identical as expected from Ref.\ \cite{Fey2015}. In comparison with the perturbation theory results, the corresponding potential curves lie energetically below the latter.

Finally, Fig.\ \ref{fig:GreenDiagCompareGD}(a) shows a result provided by the Green's function approach which cannot be obtained within the exact diagonalization scheme. In particular, in the limit $N_z \rightarrow \infty$ we obtain from Eq.\ (\ref{gf_trunc})
\begin{eqnarray}
G^{\rm (tr)}_{\rm gd}(\bfmath{R},\bfmath{R};E;\bfmath{N})=\frac{1}{2 \pi}\sum^{M_{\rm max}}_{m=-M_{\rm min}}G^{\rm (1d)}_{\rm har}(z_n,z_n;E-\varepsilon_{0m})R^{2}_{0m}(\rho_n), \label{gf_trunc_3}
\end{eqnarray}
which is inserted into Eq.\ (\ref{eq_gf_E}). Because the basis set used for exact diagonalization is necessarily finite, the result provided by Eq.\ (\ref{gf_trunc_3}) can only be analyzed in the Green's function approach. In Fig.~\ref{fig:GreenDiagCompareGD}(a), the result of this particular analysis is indicated by black dots and labeled as ``Gf-approach ($N_z \rightarrow \infty$)''. One observes that this PES lies energetically below the other curves with a maximum deviation of around $11\%$ for $\rho=0$ from the finite basis set analysis. This deviation is related to the inclusion of higher $N_z$ terms, but remains finite as the summations over the $n$ and $m$ quantum numbers are truncated at $N=0$ and $(M_{\rm min},M_{\rm max})$, respectively. Performing the unlimited summations over $n$ and $m$ would lead to a divergent shift of the potential curve as in this limit we obtain the full Green's function Eq.\ (\ref{gdgf}) (see Sec.\ \ref{gf_div}).

In Fig.\ \ref{fig:GreenDiagCompareGD}(b), we present the analogous study for the ground state PES for $\rho_n=0$ as a function of $z_n$ for the same parameters as in Fig.\ \ref{fig:GreenDiagCompareGD}(a). Because of the $z \rightarrow -z$ parity symmetry of the electronic Hamilton $H_{\rm el}$, the potential curve is symmetric in $z_n$. The potential curve along the $z_n$-direction forms a Gaussian-like well. For this reason, the three-dimensional potential surface forms a single well with a minimum at $\rho_n=z_n=0$. As we have seen in the calculations for the graphs in Fig.~\ref{fig:GreenDiagCompareGD}(a) the potential curves energetically decrease for increasing basis set. Again, the limiting case $N_{z} \rightarrow \infty$ is indicated by black dots. The deviations of these calculations with respect to the finite basis set results (solid red line) are again in the range of $11\%$. 

Finally, we derive the regularized giant-dipole Green's function $G^{\rm (reg)}_{\rm gd}$ as it was discussed in Sec.\ \ref{gf_div}. For this derivation we use a the cylindrical representation of the $1/|\bfmath{r}-\bfmath{r}^\prime|$ term which is given by \cite{Abramowitz1972}
\begin{eqnarray}
\frac{1}{|\bfmath{r}-\bfmath{r}^{\prime}|}=\frac{1}{\pi \sqrt{\rho \rho^{\prime}}}\sum^{\infty}_{m=-\infty}e^{im(\phi-\phi^\prime)}Q_{m-1/2}(\chi)
\end{eqnarray}
in terms of the Legendre-Q functions $Q_{m-1/2}(x)$ \cite{Abramowitz1972} with $\chi=(\rho^2+\rho^{\prime 2}+(z-z^{\prime})^2)/(2\rho \rho^{\prime})$. Using this relation we then easily derive the following analytic expression for the regularized giant-dipole Green's function at $\bfmath{r}=\bfmath{r}^\prime$
\begin{eqnarray}
G^{\rm (reg)}_{\rm gd}(\bfmath{r},\bfmath{r};E)=\frac{1}{2 \pi}\sum^{\infty}_{m=-\infty}\left( \sum^{\infty}_{n=0}G^{\rm (1d)}_{\text{har}}(z,z;E-\varepsilon_{nm})R^{2}_{nm}(\rho)
+\frac{1}{\pi \rho}Q_{m-1/2}(1)\right).
\end{eqnarray}
\section{Conclusion}
In this work, we have presented a Green's function analysis of giant-dipole systems. In a first step, we have derived the exact Green's function representation of several magnetically field-dressed systems. In particular, we calculated the Green's function for giant-dipole systems for arbitrary field parameters as well as in the two-dimensional harmonic approximation. Using these results of the Green's function analysis, we have calculated the dynamical polarizability. This property is of special interest as it characterizes the systems response to an external time-dependent driving field, which provides insights into the dynamics of interacting atom-laser systems. Furthermore, we have applied the Green's function approach to deduce electronic properties for a specific species of ultra-long range molecules, so-called diatomic giant-dipole molecules. In particular, we have calculated the adiabatic potential energy surfaces which we compare to results that have been obtained within an exact diagonalization approach \cite{Dippel1994,Kurz12}. Beside the specific electronic potential curves of these particular molecular species, we have also shown a general divergent behavior in the potential energy calculation of ultra-long range molecules within the Fermi-pseudopotential approach, which indicates a general insufficiency of this specific ansatz. For this reason, we derived the regularized Green's function of the giant-dipole problem, as this approach has proven to avoid the insufficiencies of the pseudopotential approach in previous works \cite{Fabrikant2002,Fey2015}.

Although our study has provided novel information on the Green's function approach of giant-dipole systems, there are still open questions. For instance, a derivation of a closed representation of the giant-dipole Green's function would be useful for the implementation of numerically robust routines. Furthermore, in this work we have only considered a pure triplet pseudopotential for the electron-perturber interaction. Thus, the question arises how the inclusion of both the singlet scattering channel as well a hyperfine structure of the perturber atoms change the PES. Furthermore, the inclusion of higher-order terms in the Fermi-pseudopotential ansatz has shown to provide additional features for both single- and multi-perturber ultra-long range molecules \cite{Hamilton2002,Fey2016}. In particular, in Ref.\ \cite{Fey2016} it was shown that the Green's function ansatz is quite suitable to study multi-perturber systems, which is an open question for molecular giant-dipole systems as well. For this reason, molecular giant-dipole systems provide a plethora of interesting problems which can be addressed in future studies.
\section*{Acknowledgments}
T.\ S.\ acknowledges financial support from ``Evangelisches Studienwerk Villigst''. 
\section*{Appendix A (Derivation of the Green's function expression)}
We show that Eq.\ (\ref{GFtotalexp}) solves Eq.\ (\ref{GFtotal}) by performing the corresponding calculation. We get
\begin{eqnarray}
&\ &(E-H)G(\bfmath{r},\bfmath{r}';E) \nonumber\\
&=&\sum_{k_2...k_N}(E-H)G_1(\bfmath{r},\bfmath{r}^{\prime}_{1};E-\sum^{N}_{i=2}\varepsilon^{(i)}_{k_{i}})\prod^{N}_{n=2}\phi^{(n)*}_{k_{n}}(\bfmath{r}^{\prime}_{n}) \phi^{(n)}_{k_{n}}(\bfmath{r}_{n})\ \ , \ \ \ \ \ \left( H=\sum^{N}_{i=1}h_i \right) \nonumber\\
&=&\sum_{k_2...k_N}[ E-\sum^{N}_{i=2}\varepsilon^{(i)}_{k_{i}}-h_1+\sum^{N}_{i=2}(\varepsilon^{(i)}_{k_{i}}-h_i) ] G_1(\bfmath{r},\bfmath{r}^{\prime}_{1};E-\sum^{N}_{i=2}\varepsilon^{(i)}_{k_{i}})\prod^{N}_{n=2}\phi^{(n)*}_{k_{n}}(\bfmath{r}^{\prime}_{n}) \phi^{(n)}_{k_{n}}(\bfmath{r}_{n}) \nonumber\\
&=&\sum_{k_2...k_N}(E-\sum^{N}_{i=2}\varepsilon^{(i)}_{k_{i}}-h_1)G_1(\bfmath{r},\bfmath{r}^{\prime}_{1};E-\sum^{N}_{i=2}\varepsilon^{(i)}_{k_{i}})\prod^{N}_{n=2}\phi^{(n)*}_{k_{n}}(\bfmath{r}^{\prime}_{n}) \phi^{(n)}_{k_{n}}(\bfmath{r}_{n}) \nonumber\\
&\ &+\sum_{k_2...k_N}\sum^{N}_{i=2}(\varepsilon^{(i)}_{k_{i}}-h_i))G_1(\bfmath{r},\bfmath{r}^{\prime}_{1};E-\sum^{N}_{i=2}\varepsilon^{(i)}_{k_{i}})\prod^{N}_{n=2}\phi^{(n)*}_{k_{n}}(\bfmath{r}^{\prime}_{n}) \phi^{(n)}_{k_{n}}(\bfmath{r}_{n}). \label{derivation}
\end{eqnarray}
Using Eqs.\ ((\ref{eigen}),(\ref{delta})) we get
\begin{eqnarray*}
(E-\sum^{N}_{i=2}\varepsilon^{(i)}_{k_{i}}-h_1)G_1(\bfmath{r},\bfmath{r}^{\prime}_{1};E-\sum^{N}_{i=2}\varepsilon^{(i)}_{k_{i}})=\delta(\bfmath{r}_1-\bfmath{r}^{\prime}_{1}),\ \ \textrm{and}\ \ h_i \phi^{(i)}_{k_{i}}(\bfmath{r}_i) = \varepsilon^{(i)}_{k_i}\phi^{(i)}_{k_{i}}(\bfmath{r}_{i}).
\end{eqnarray*}
For this reason the second term in Eq.\ (\ref{derivation}) cancels and the first term can be rewritten to
\begin{eqnarray*}
 \sum_{k_2...k_N}\delta(\bfmath{r}_1-\bfmath{r}^{\prime}_{1})\prod^{N}_{n=2}\phi^{(n)*}_{k_{n}}(\bfmath{r}^{\prime}_{n}) \phi^{(n)}_{k_{n}}(\bfmath{r}_{n})=\prod^{N}_{n=1}\delta(\bfmath{r}_n-\bfmath{r}^{\prime}_{n})=\delta(\bfmath{r}-\bfmath{r}^{\prime}),
\end{eqnarray*}
which is the expression that is to be derived.\\
\\
Alternatively, one can derive Eq.\ (\ref{GFtotalexp}) by using Eq.\ (\ref{GFeigenexp}). Here, one has to consider the fact that the total wave function $\psi$ is a product state of the $\phi$-functions and that the total eigenenergies are sums of the single Hamiltonian eigenenergies $\varepsilon^{(i)}_{k}$. We get
\begin{eqnarray*}
G(\bfmath{r},\bfmath{r}^{\prime};E)&=&\sum_{k_1...k_N}\frac{\prod^{N}_{n=1}\phi^{(n)*}_{k_{n}}(\bfmath{r}^{\prime}_{n}) \phi^{(n)}_{k_{n}}(\bfmath{r}_{n})}{E-\sum^{N}_{i=1}\varepsilon^{(i)}_{k_{i}}}\\
&=& \sum_{k_2...k_N}\sum_{k_1}\frac{\phi^{(n)*}_{k_1}(\bfmath{r}^{\prime}_{1}) \phi^{(n)}_{k_1}(\bfmath{r}_1)}{E-\sum^{N}_{i=2}\varepsilon^{(i)}_{k_{i}}-\varepsilon_1}\prod^{N}_{n=2}\phi^{(n)*}_{k_{n}}(\bfmath{r}^{\prime}_{n}) \phi^{(n)}_{k_{n}}(\bfmath{r}_{n})\\
&=&\sum_{k_2...k_N}G_{1}(\bfmath{r},\bfmath{r}^{\prime}_{1};E-\sum^{N}_{i=2}\varepsilon^{(i)}_{k_{i}})\prod^{N}_{n=2}\phi^{(n)*}_{k_{n}}(\bfmath{r}^{\prime}_{n}) \phi^{(n)}_{k_{n}}(\bfmath{r}_{n}).
\end{eqnarray*}
In the final step we used the expansion of the $G_1$ Green's function (see Eq.\ (\ref{GFeigenexp})). Obviously, any Green's function $G_i$ can be used to express the total function $G(\bfmath{r},\bfmath{r}^{\prime};E)$. 
\section*{Appendix B (Calculation of giant-dipole polarizabilities)}
The giant-dipole system for arbitrary field strengths can be transformed via a unitary transformation into a systems of three decoupled harmonic oscillators with Green's function $\tilde{G}^{\rm (3d)}_{\rm har}(\bfmath{r},\bfmath{r}^\prime;E)$ \cite{Dippel1994}. Because the eigenfunctions $\Psi_n(\bfmath{r})$ and the giant-dipole Green's function transform according to
\begin{eqnarray}
| \Psi_n \rangle=U|\phi_{n_1}\phi_{n_2}\phi_{n_z}\rangle,\ \ \ \mathcal{G}_{\rm gd}(E)=U\tilde{G}^{\rm (3d)}_{\rm har}(E)U^{\dagger}
\end{eqnarray}
we get
\begin{eqnarray}
\alpha^{(n)}_{ii}(\omega)&=&-\sum_{\sigma = \pm 1}\langle \phi_{n_1}\phi_{n_2}\phi_{n_z} |U^{\dagger}x_iU\tilde{G}^{\rm (3d)}_{\rm har}(\omega_n \pm \sigma \omega)U^{\dagger}x_iU|\phi_{n_1}\phi_{n_2}\phi_{n_z}\rangle
\end{eqnarray}
The transformation of the spatial coordinates under the unitary transformation is explicitly given as \cite{Meyer1988}
\begin{eqnarray*}
 U^{\dagger}xU=x-\beta p_y,\ \ \ U^{\dagger}yU=y-\beta p_x,\ \ \ U^{\dagger}zU=z,
\end{eqnarray*}
which leads to the expressions for the dynamical polarizabilities
\begin{eqnarray*}
\alpha^{(n)}_{xx}(\omega)&=& -\sum_{\sigma = \pm 1}\langle \phi_{n_1}\phi_{n_2}\phi_{n_z} |(x-\beta p_y)\tilde{G}^{\rm (3d)}_{\rm har}(\omega_n \pm \sigma \omega)(x-\beta p_y)|\phi_{n_1}\phi_{n_2}\phi_{n_z}\rangle,\\
\alpha^{(n)}_{yy}(\omega)&=& -\sum_{\sigma = \pm 1}\langle \phi_{n_1}\phi_{n_2}\phi_{n_z} |(y-\beta p_x)\tilde{G}^{\rm (3d)}_{\rm har}(\omega_n \pm \sigma \omega)(y-\beta p_x)|\phi_{n_1}\phi_{n_2}\phi_{n_z}\rangle,\\
\alpha^{(n)}_{zz}(\omega)&=& -\sum_{\sigma = \pm 1}\langle \phi_{n_1}\phi_{n_2}\phi_{n_z} |z\tilde{G}^{\rm (3d)}_{\rm har}(\omega_n \pm \sigma \omega)z|\phi_{n_1}\phi_{n_2}\phi_{n_z}\rangle.
\end{eqnarray*}
Obviously, the result of $\alpha^{(n)}_{zz}(\omega)$ remains unchanged and is given by Eq.\ (\ref{pol_dyn}), i.e.\ $\alpha^{(n)}_{zz}(\omega)=\alpha^{||}_{n}(\omega)$.
\section*{Appendix C (Regularized Green's function)}
Analogous to Ref.\ \cite{Fey2015}, we define the regularized Green's function $\tilde{G}^{\rm (reg)}_{0}$ as
\begin{eqnarray*}
\tilde{G}^{\rm (reg)}_{0}(\bfmath{r},\bfmath{r}^\prime;E) \equiv \frac{\partial}{\partial \xi}\left(\xi G_{0}(\bfmath{r},\bfmath{r}^\prime;E) \right),\ \ \ \xi=|\bfmath{r}-\bfmath{r}^\prime|.
\end{eqnarray*}
With $G_{0}=-1/(2\pi \xi)+G^{\rm (reg)}_{0}$ we obtain
\begin{eqnarray*}
\tilde{G}^{\rm (reg)}_{0}(\bfmath{r},\bfmath{r}^\prime;E)= G_{0}(\bfmath{r},\bfmath{r}^\prime;E)+\frac{1}{2\pi \xi}+\xi \left( \frac{\partial}{\partial \xi}G^{\rm (reg)}_{0}(\bfmath{r},\bfmath{r}^\prime;E) \right).
\end{eqnarray*}
In the limit $\xi \rightarrow 0$ we then find
\begin{eqnarray*}
\tilde{G}^{\rm (reg)}_{0}(\bfmath{r},\bfmath{r};E)=\lim_{\xi \rightarrow 0}\left( G_{0}(\bfmath{r},\bfmath{r}^\prime;E)+\frac{1}{2\pi \xi} \right) + \underbrace{\lim_{\xi \rightarrow 0}\xi \left( \frac{\partial}{\partial \xi}G^{\rm (reg)}_{0}(\bfmath{r},\bfmath{r}^\prime;E) \right)}_{=0}=G^{\rm (reg)}_{0}(\bfmath{r},\bfmath{r};E).
\end{eqnarray*}
\section*{Appendix D (Oscillations of the truncated giant-dipole Green's function)}
\begin{figure}
\centering
\includegraphics[width=0.6\textwidth]{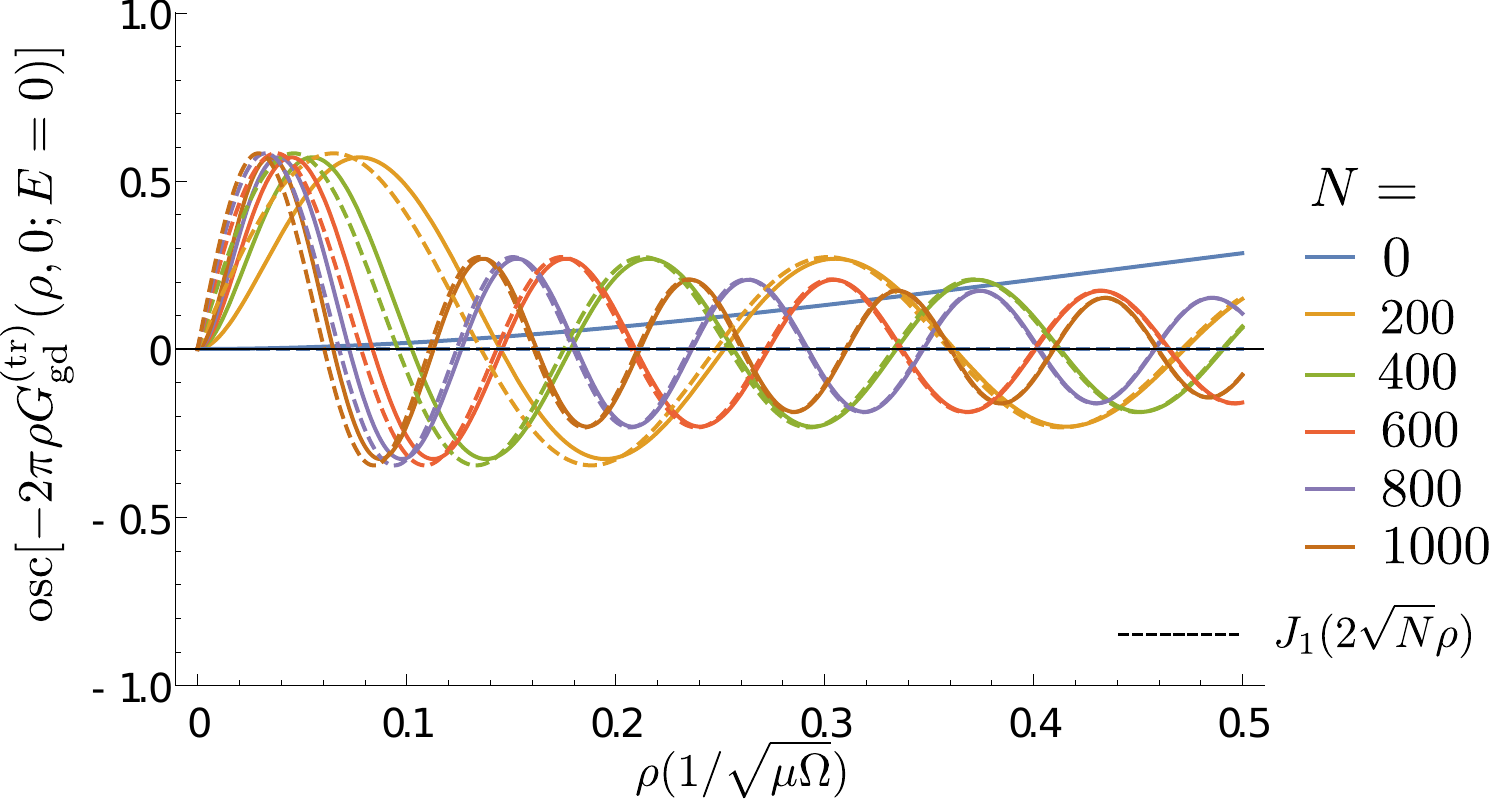} 
\caption{Oscillatory part $\textrm{osc}[-2\pi \rho G^{(\rm tr)}_{\rm gd}(\rho,0;E=0)]$ (solid curves) in comparison to a Bessel function approximation given by $J_{1}(2 \sqrt{N}\rho)$ (dashed curves).}
\label{fig:osci}
\end{figure}

The oscillatory part of $G^{(\rm tr)}_{\rm gd}(\rho,0;E=0)$ 
\begin{eqnarray*}
\textrm{osc}[-2\pi \rho G^{(\rm tr)}_{\rm gd}(\rho,0;E=0)]=2 \pi \rho \left(G^{(\rm tr)}_{\rm gd}(z=\rho,0;E=0)-G^{(\rm tr)}_{\rm gd}(\rho,0,E=0)\right) \approx J_{1}(2 \sqrt{N}\rho)
\end{eqnarray*}
for $\bfmath{r}\rightarrow 0$ is shown in Fig.\ \ref{fig:osci} dependent on the summation limit $N$. For sufficiently large $N$ it is well approximated by a $J_1$-Bessel function \cite{Abramowitz1972}, in particular $J_{1}(2 \sqrt{N}\rho)$. In the case of $N \rightarrow \infty$ it is $J_{1}(2 \sqrt{N}\rho)\rightarrow 0$ for $\rho > 0$. For this reason the oscillations vanish in the limit of the full giant-dipole Green's function.
\section*{Bibliography}

\begin{thebibliography}{0}
\bibitem{Baye1992}Baye D, Clerbaux N and Vincke M 1992 {\it Phys.\ Lett.\ A} {\bf 166} 135 
\bibitem{Dzyaloshinskii1992}Dzyaloshinskii I 1992 {\it Phys.\ Lett.\ A} {\bf 165} 69
\bibitem{Dippel1994}Dippel O, Schmelcher P and Cederbaum L S 1994 {\it Phys. Rev. A} {\bf 49} 4415 
\bibitem{Schmelcher1998}Shertzer J, Ackermann J and Schmelcher P 1998 {\it Phys.\ Rev.\ A} {\bf 58} 1129
\bibitem{Schmelcher2001}Schmelcher P 2001 {\it Phys.\ Rev.\ A} {\bf 64} 063412 
\bibitem{Fauth1987}Fauth M, Walther H and Werner E 1987 {\it Z.\ Phys.\ D} {\bf 7} 293
\bibitem{Raithel1993}Raithel G, Fauth M and Walther H 1993 {\it Phys.\ Rev.\ A} {\bf 47} 419 
\bibitem{Schmelcher1993a}Schmelcher P and Cederbaum L S 1993 {\it Chem.\ Phys.\ Lett.} {\bf 208} 548 
\bibitem{Avron1978}Avron J E, Herbst I W and Simon B 1978 {\it Ann.\ Phys.\ (NY)} {\bf 114} 431
\bibitem{Herold1981}Herold H, Ruder H and Wunner G 1981 {\it J.\ Phys.\ B} {\bf 14} 751
\bibitem{Johnson1983}Johnson B r, Hirschfelder J O and Yang K H {\it Rev.\ Mod.\ Phys.} {\bf 55} 109 
\bibitem{Ackermann1997}Ackermann J Shertzer J and Schmelcher P 1997 {\it Phys.\ Rev.\ Lett.} {\bf 78}  199
\bibitem{Shertzer1998}Shertzer J, Ackermann J and Schmelcher P 1998 {\it Phys.\ Rev.\ A} {\bf 58} 1129.
\bibitem{Kurz2017}Kurz M, Gr\"unwald P, and Scheel S 2017 {\it Phys. Rev. B} {\bf 95} 245205 
\bibitem{Kurz12} Kurz M, Mayle M and Schmelcher P 2012 {\it Europhys. Lett.} {\bf{97}} 43001 
%
\bibitem{Greene2001}Greene C H, Dickinson A S and Sadeghpour H R 2000 {\it Phys. Rev. Lett.} {\bf 85} 2458
\bibitem{Hamilton2002}Hamilton E L, Greene C H and Sadeghpour H R 2002 {\it J. Phys. B: At. Mol. Opt. Phys.} {\bf 35} L199–L206
\bibitem{Stanojevic2006}Stanojevic J, C\^{o}t\'{e} R, Tong D, Farooqi S M, Eyler E E, and Gould P L 2006 {\it Eur.\ Phys.\ J.\ D} {\bf 40} 3–12 
\bibitem{Bendkowsky2009}Bendkowsky V, Butscher B, Nipper J, Shaffer J P, L\"ow R and Pfau T 2009 {\it Nature} { \bf 458} 1005
\bibitem{Tallant2012}Tallant J,  Rittenhouse S T, Booth D, Sadeghpour H R, and Shaffer P 2012 {\it Phys.\ Rev.\ Lett.} {\bf 109} 173202 
\bibitem{Krupp2014}Krupp A T {\it et al.} 2014 {\it Phys. Rev. Lett.} {\bf 112} 143008
\bibitem{Gonzalez2014}Gonz\'{a}les-F\'{e}rez R, Sadeghpour H R, and Schmelcher P 2017 {\it New J.\ Phys.} {\bf 17} 013021
\bibitem{Anderson2014}Anderson D A, Miller S A and Raithel G 2014 {\it Phys. Rev. Lett.} {\bf 112} 163201
\bibitem{Booth2015}Booth D, Rittenhouse S T, Yang J, Sadeghpour H R, Shaffer J P 2015 {\it Science} {\bf 348} 99-102
\bibitem{DeSalvo2015}DeSalvo B J, Aman J A, Dunning F B, Killian TC, Sadeghpour H R, Yoshida S, and Burgd\"orfer J 2015 {\it Phys.\ Rev.\ A} {\bf 92} 031403(R)
\bibitem{Sassmannshausen2015}Sa\ss{}mannshausen H, Merkt F, and Deiglmayr J 2015 {\it Phys.\ Rev.\ Lett.} {\bf 114}  133201
\bibitem{Sassmannshausen2016}Sa\ss{}mannshausen H, Deiglmayr J, and Merkt F 2016 {\it Eur. Phys. J. Special Topics} {\bf 225} 2891-2918
\bibitem{Niederpruem2016}Niederpr\"um T, Thomas O, Eichert T, and Ott H 2016 {\it Phys.\ Rev.\ Lett.} {\bf 117} 123002
\bibitem{Niederpruem2016b}Niederpr\"um T, Thomas O, Eichert T, Lippe C, P\'{e}rez-R\'{i}oz J, Greene C H, and Ott H 2016 {\it Nature Communications} {\bf 7} 12820
\bibitem{Camargo2016}Camargo F, Whalen J D, Ding R, Sadeghpour H R, Yoshida S, Burgd\"orfer J, Dunning F B, and Killian T C 2016 {\it Phys.\ Rev.\ A} {\bf 93} 022702
\bibitem{Gonzalez2016}Fernandez J A, Schmelcher P, and Gonz\'{a}les-F\'{e}rez R 2016 {\it J.\ Phys.\ B: At.\ Mol. Opt.\ Phys.} {\bf 49} 124002
%
\bibitem{Fermi1934}Fermi E 1934 {\it Nuovo Cimento} {\bf 11} 157
\bibitem{Fabrikant2002}Khuskivadze A A, Chibisov M I, Fabrikant I I 2002 {\it Phys.\ Rev.\ A} {\bf 66}, 042709
\bibitem{Greene2006}Greene C H, Hamilton E L, Crowell H, Vadla C, and Niemax K 2006 {\it Phys.\ Rev. Lett.} {\bf 97} 233002
\bibitem{Bendkowsky2010} Bendkowsky V {\it et al.} 2010 {\it Phys.\ Rev.\ Lett.} {\bf 105} 163201
\bibitem{Fey2015}Fey C, Kurz M, Schmelcher P, Rittenhouse S T and Sadeghpour H 2015 {\it New J. Phys.} {\bf 17} (2015) 055010
\bibitem{Fey2016}Fey C, Kurz M and Schmelcher P 2016 {\it Phys.\ Rev.\ A} {\bf 94} 012516
\bibitem{atomicunits}If not stated otherwise atomic units are used throughout this work.
\bibitem{Bychov1960}Bychov Y A 1960 {\it J.\ Exptl.\ Theoret.\ Phys.\ (U.S.S.R.)} {\bf 39} 689
\bibitem{Ueta1992}Ueta T 1992 {\it J.\ of the Phys.\ Soc.\ of Jap.} {\bf 61} 4314
\bibitem{Abramowitz1972}Abramowitz M and Stegun I A (Eds.) 1972 {\it Handbook of Mathematical Functions with Formulas, Graphs, and Mathematical Tables} 9th printing.\ New York: Dover
\bibitem{Bakhrakh1970}Bakhrakh V L, Vetchinkin S I 1970 {\it Theor Math Phys} {\bf 6} 392
\bibitem{Bakhrakh1972}Bakhrakh V L, Vetchinkin S I and Khristenko S V 1972 {\it Theor Math Phys} {\bf 12} 223
\bibitem{Davydkin1971}Davydkin V A, Zon B A, Manakov N L and Rapoport L P 1971 {\it Soviet Physics JETP} {\bf 33} 70
\bibitem{Omont1977} Omont A 1977 {\it J. Phys. (Paris)} {\bf 38} 1343
\bibitem{Eiles2016}Eiles M T, P\'{e}rez-R\'{i}oz J, Robicheaux F and Greene C H 2016 {\it J. Phys. B: At. Mol. Opt. Phys.} {\bf 49} 114005 
\bibitem{Meyer1988}Meyer H D, Kucar J and Cederbaum L S 1988 {\it Journal of Mathematical Physics} {\bf 29} 1417 
\end{thebibliography}
\end{document}